\documentclass{aa}
\usepackage{txfonts}
\usepackage{graphicx}
\usepackage{siunitx}
\usepackage{amsmath,amssymb}
\usepackage{multirow}
\usepackage{comment}
\usepackage{hyperref}

\makeatletter
\renewcommand*\aa@pageof{, page \thepage{} of \pageref*{LastPage}}
\makeatother
              
\begin{document}

\title{A transformer-based generative model for planetary systems}

\author{Yann Alibert \inst{1,2} \and Jeanne Davoult \inst{1,3} \and Sara Marques \inst{1,2}\fnmsep\thanks{Code available at \url{ai4e.eu} and \url{ai4exoplanets.com}}}

\institute{Space Research and Planetary Sciences, Physics Institute, University of Bern, Gesellschaftsstrasse 6, 3012 Bern, Switzerland
\and
Center for Space and Habitability, University of Bern, Gesellschaftsstrasse 6, 3012 Bern, Switzerland
\and
Institut für Planetenforschung, German Aerospace Center (DLR), Rutherfordstrasse 2, 12489 Berlin, Germany
}
\date{\today} 

\abstract{Numerical calculations of planetary system formation are very demanding in terms of computing power. These synthetic planetary systems can however provide access to correlations, as predicted in a given numerical framework, between the properties of planets in the same system. Such correlations can, in return, be used in order to guide and prioritize observational campaigns aiming at discovering some types of planets, as Earth-like planets.}
{Our goal is to develop a generative model which is capable of capturing correlations and statistical relationships between planets in the same system. Such a model, trained on the Bern model, offers the possibility to generate large number of synthetic planetary systems with little computational cost, that can be used, for example, to guide observational campaigns.}
{We use a training database of ~25000 planetary systems with up to 20 planets and assuming a solar-type star, generated using the Bern model. Our generative model is based on the transformer architecture which is well-known to efficiently capture correlations in sequences and is at the basis of all modern Large Language Models. To assess the validity of the generative model, we perform visual and statistical comparisons, as well as a machine learning driven tests. Finally, as a use case example, we consider the TOI-469 system, in which we aim at predicting the possible properties of planets c and d, based on the properties of planet b (the first that has been detected).}
{We show using different comparison methods that the properties of systems generated by our model are very similar to the ones of the systems computed directly by the Bern model. We also show that different classifiers cannot distinguish between the directly computed and generated populations, adding confidence that the statistical correlations between planets in the same system are similar. Lastly, we show in the case of the TOI-469 system, that using the generative model allows to predict the properties of planets not yet observed, based on the properties of the already observed planet.}
{Our generative model, which we provide to the community on our \href{www.ai4exoplanets.com}{website}, can be used to study a variety of problems like understanding correlations between certain properties of planets in systems, or predicting the composition of a planetary system, given some partial information (e.g. presence of some easier-to-observe planets). Nevertheless, it is important to note that the performances of our generative model rely on the ability of the underlying numerical model, here the Bern model, to accurately represent the actual formation process of planetary system. Our generative model could, on the other hand, very easily be re-trained using, as input, other numerical models provided by the community.}


\authorrunning{Alibert, Davoult and Marques}

\maketitle 

\section{Introduction}
\label{introduction}

Major observational projects are presently undergoing or being studied in order to detect and characterize low mass planets in the habitable zone of their star. One can cite for example the ESPRESSO spectrograph \citep{Pepe_etal_2021}, which is precise enough to detect the radial-velocity effect of planets similar in mass and period to the Earth, the CHEOPS space telescope \citep{Benz_etal_2021} which can detect also the transit of these kinds of planets, the future the LIFE mission concept \citep{paper_LIFE_1,paper_LIFE_2,paper_LIFE_3,paper_LIFE_4}, that aims at observing in the near-IR planets orbiting our close neighbors, or the Habitable World Observatory (HWO), which, as the name indicates, should be able to detect and characterize, in the visible and UV domain, planets like the Earth. For many of these facilities, however, the detection and/or observation of low-mass, long-period planets, while possible, is extremely time consuming, and blind searches can make it impossible. 

To avoid blind searches, any observable properties of planetary systems that could indicate a high (or low) likelihood to harbor a planet like ours are useful. Such properties could be the ones of the central star (mass, composition, age, etc...), of the stellar environment (e.g. the presence of a close companion could hinder the presence of any planet in the habitable zone), or of other planets already observed in the same system. As a simple example, the presence of a Jupiter-like planet within 1 \unit{\astronomicalunit} of its central star, in the case of a solar type star, is a strong indication that such planetary system is not the best for search an low-mass planet in the habitable zone, from the point of view of dynamical stability, \citep{Latham2011, Steffen2012}).

In the recent years, several advancements in observation and theory have revealed that planetary systems often exhibit distinct architectural patterns. One well-known example is the peas-in-a-pod configuration \citep{Millholland_etal_2017,Weiss_etal_2018,He_etal_2019,PP7_Weiss}. In this configuration, planets within the same system tend to have similar radii and closely spaced orbits, as indicated by their period ratios (though some papers have questioned this trend, see e.g. \citep{Zhu_2020} and \citep{Murchikova_Tremaine_2020}). Additionally, different studies based on different types of planetary system formation models have also studied the architecture of planetary systems from a theoretical point of view \citep{Mishra_2023a,Mishra_2023b,Emsenhuber_etal_2023}. Another example includes the information theory approach proposed by \citep{Gilbert2020}, in which planetary systems patterns and classification are addressed in a more descriptive approach in order to avoid the bias from physical assumptions.

Notwithstanding, such architecture patterns can be useful for different reasons. First, the architecture of planetary systems, its relationship with stellar metallicity (e.g., \citep{Brewer2018, Zhu&Wu2018,Ghezzi2021,Zhu2024,Bryan2024}), and planetary internal structure and composition offer unique and innovative insights into understanding the formation, migration and evolution of multi-planet systems 
\citep{Winn_Fabrycky_2015}. Secondly, the  architecture of planetary systems can be used to predict the presence and properties of planets not presently detected. Based on this idea and inspired by the architecture classification framework of \citep{Mishra_2023a,Mishra_2023b}, \citep{Davoult_etal_2024} have computed the probability for a given planetary system to harbor an Earth-like planet. They showed that there is a correlation between the architecture class of a system, and the presence of an Earth-like planet. Similarly, \citep{Davoult_etal_2024} demonstrated that the properties of the inner observable planet in a system can be used to predict the presence of an Earth-like planet in the same system. 

In a more general way, the above-mentioned observational facilities would tremendously benefit from the possibility to estimate, in the most general way, the probability of presence of some type of planet in a system, given some observed properties of other planets (and star) in the same system. Such a problem means being able to estimate the conditional probabilities of the presence and properties of some planets in a system, given a set of observations of that same system.

Such conditional properties can be easily estimated using a planetary system formation model that would compute the final properties of synthetic planetary systems. This, however, requires that the numerical model can be ran millions of times, in order to, for example, down-select in a multi-million-size database of synthetic planetary systems, i.e, the sub-part which matches some observed properties to finally compute which fraction of those harbor an Earth-like planet (if the goal is to find such a planet). Unfortunately, planetary system formation models rely on costly simulations solving ensembles of differential equations for different sets of initial conditions
\citep{PP7_Drajkowska,NGPPS1,NGPPS2}. This very high computational cost (typically weeks on a single core for one single planetary system) prevents computing massive populations, which are required to predict the properties of unknown planets in a system given a limited set of observations of the same system.

A first effort to classify planetary systems from an observational point of view using machine learning and a linguistic framework is described in \citep{Sandford2021}. While they perform better than a naive approach and are able to find correlations in order to classify planets, host stars and planetary systems, they suffer from the observational bias.
Additionally, \citep{DYNAMITE2020, DYNAMITE2024} provides a modular framework based on population statistics used to be able to predict undetected planets with some observational examples \citep{DYNAMITE2020a, DYNAMITE2022, Basant2022}. While this work is also very insightful, it relies on the assumption that the properties defining the planets and their orbits are statistically independent.

In this paper, we train a planetary system generative model, to emulate the results of a population synthesis model, in this case, the Bern model \citep{NGPPS1,NGPPS2,NGPPS3,NGPPS4,NGPPS5,NGPPS6}, and generate millions of systems in only a few minutes. As such, our approach is not subject to observational bias, and can generate complete planetary systems. Additionally, we do not need to make further assumptions other than the ones in the Bern model. Finally, we note that the approach we develop here can be used with any planetary system formation model predicting the planetary properties used to train the model (mass and semi-major axis in this paper).

The paper is organized as follows: in Section \ref{NGPPS} we present the planetary system formation models that are used to train our generative model. In Section \ref{generative_model}, we present the architecture of our generative model, as well as the training procedure. In Section \ref{results}, we compare the results provided by our generative model with the ones from the direct numerical simulations of Section \ref{NGPPS}. Finally, Section \ref{conclusion} is devoted to discussion and conclusion. 

\section{Planetary system formation model}
\label{NGPPS}

Our generative model is trained on a database of synthetic planetary systems. Before describing the database we used, it is very important to emphasize the fact that the prediction of our generative model will, by construction, completely depend on the database that was used to train it. In other words, using our generative model to estimate conditional probabilities, as described above, implicitly assumes that the numerical model used to generate the database reflects the actual formation process of real planetary systems.

In this paper, we use results of Next Generation Planetary Population Synthesis (NGPPS) computed using the Bern model \citep{NGPPS1,NGPPS2,NGPPS3,NGPPS4,NGPPS5,NGPPS6} to generate a database of ~25000 systems orbiting solar-type stars. We summarize in the following lines the main features of the Bern model.

The Bern model is a global model of planetary system formation that uses the population synthesis method, as described in detail in \citep{Mordasini2018}. The model encompasses dozens of physical processes from a system's initial state to its final state, 10 Gyr later. 
In the model, planets form by core accretion \citep{Pollack1996} and accrete in the oligarchic regime (e.g., \citep{Ida&Makino1993, Ohtsuki2002, Fortier_etal_2013}). The formation phase lasts 20~Myr, during which planetary embryos embedded in a disc of gas and dust accrete from the material to form the final planets, but also migrate and interact dynamically thanks to an N-body simulator (ejections, giant impacts, etc). The gas disc disappears under the effect of photo-evaporation. After 20~Myr, the model switches to the evolution phase, following the planets' thermodynamic evolution (cooling, contracting, D-burning if necessary and atmospheric escape) until 10~Gyr. The model and the parameters used are detailed in \citep{NGPPS1, NGPPS2}.
A population synthesis is used to produce a population of synthetic planetary systems based on the same model, but which vary according to the initial conditions. These initial conditions are drawn from statistical distributions constrained by observations. In the same population, some parameters remain fixed (the mass of the central star, the number of planetary embryos, the gas viscosity, and the distribution of gas and planetesimals in the protoplanetary disc \citep{Veras2004}, the size of the planetesimals, and their density). The initial conditions which are drawn at random from statistical distributions are: the mass of the gas disc \citep{Beckwith+1996}, the external photo-evaporation rate M$_{wind}$ \citep{Haisch2001}, the dust-to-gas ratio, f$_{D/G}$ \citep{Murray2001,Santos2003}, the inner edge of the gas disc, R$_{in}$, and the initial location of the embryos.

The population used in this study comprises 24365 systems, generated using 20 planetary embryos, all of which after the formation and evolution phases have between one and 20 planets.

\section{Generative model}
\label{generative_model}

\subsection{Large Language Model and planetary system generation}
\label{encoding}

Large Language Models (LLMs) are now part of our everyday life and a lot of theoretical effort has been spent to increase their performances. In essence, a LLM (in its generative form) is a model that is able to predict a word given all the other words preceding it. It is therefore a system that computes the conditional probability of tokens (words or parts of words), given all the previous tokens. 

The generation of planetary systems can be framed in a similar way, where `tokens' represent synthetic planets, characterized by some of their physical properties. Generating a planetary system then relies on computing the conditional probability of planet $n+1$ given the properties of planets $1$ to $n$. In this paper, we will characterize planets only by their mass and semi-major axis, but the extension to other cases (planets being characterized by their radius, composition, etc...) can be done using a similar approach.

\begin{figure}
    \centering
    \includegraphics[width=\linewidth]{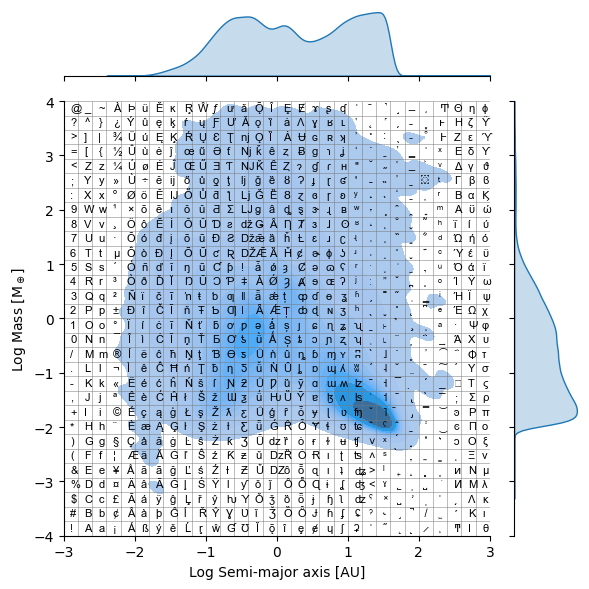}
    \caption{Example of the uniform encoding grid of $N^2$ rectangles (with $N=30$) used to encode the planets over the logarithm of the mass and semi-major axis of the planets. For each rectangle, we attribute a unique Unicode character which we use to encode a planet which is lying inside. In blue is shown the density distribution for the training population of ~25000 planetary systems used to train the model. The levels correspond to fractions of 0.0001, 0.001, 0.01, 0.1, 0.2, 0.3, 0.4, 0.5, 0.6, 0.7, 0.8, 0.9 and 1. Note that if two planets (from the same system) are located in the same rectangle, they will be encoded as two identical characters.}
    \label{fig:encoding}
\end{figure}

\begin{figure}[htp!]
    \centering
    \includegraphics[width=0.95\linewidth]{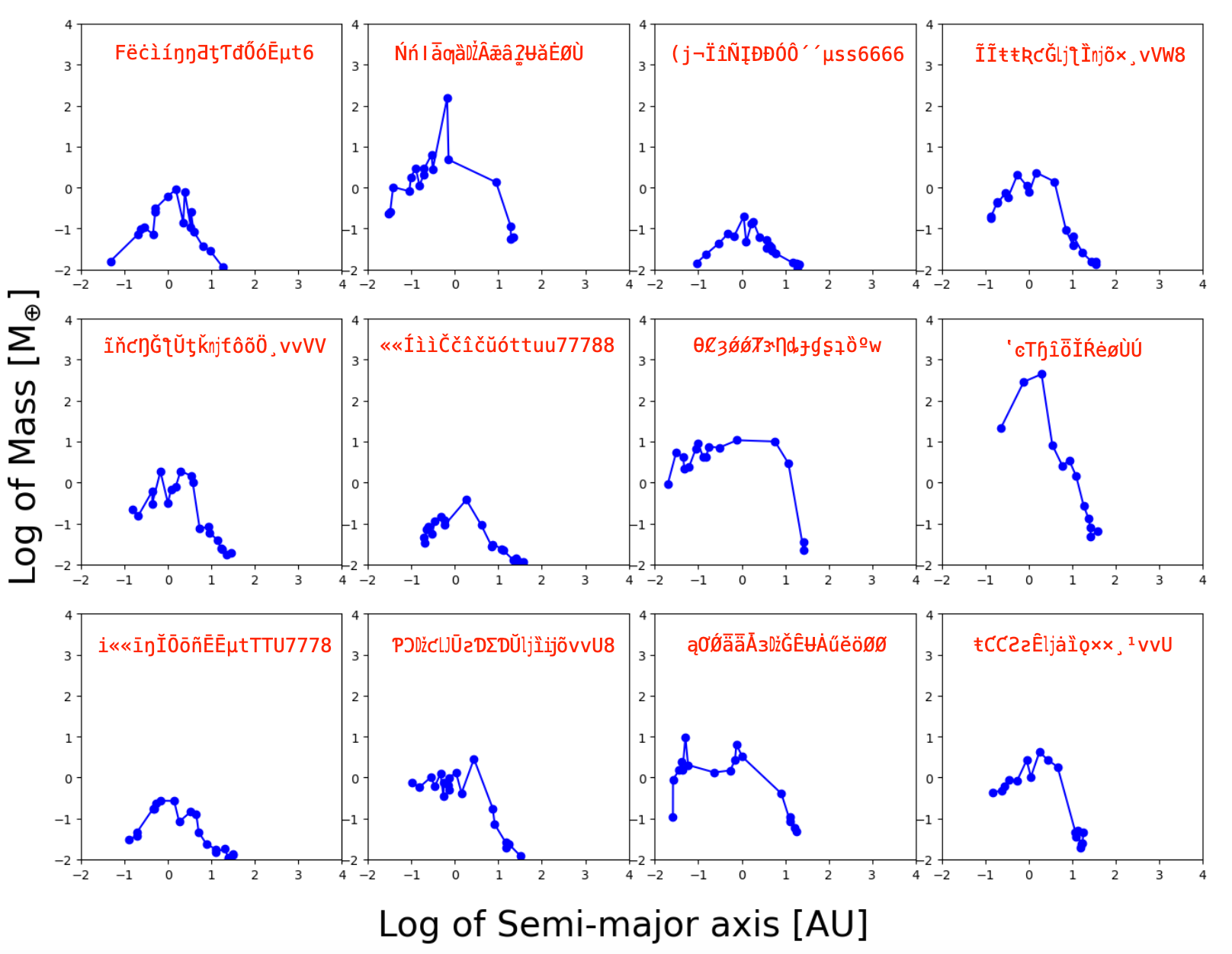}
    \caption{Example of some planetary systems belonging to our training database. The x axis represents the log of semi-major axis (in AU), the y axis represents the log of mass (in Earth masses). Each planetary system is represented as a broken line joining points, themselves representing the planets. The characters in red in each of the panels correspond to the encoding of the planetary system into a word (see Sect. \ref{encoding}).}
    \label{fig:real_PS}
\end{figure}

We developed our generative model following this procedure:
\begin{itemize}
    \item each planetary system computed by our numerical model (see \ref{NGPPS}) is represented as a word made of characters, where each character represents a planet. Since our training data consists of planetary systems with up to 20 planets, each system is represented as a word made with up to 20 characters. Each planetary system is, firstly, ordered in increasing semi-major axis, transforming the unordered set representing the collection of planets in a given system to an ordered sequence \citep{Sandford_etal_2021}.
    \item the correspondence between a planet and a character is done simply mapping the position of a planet in a mass \textit{versus} semi-major axis diagram (where the x axis represents the log of semi-major axis in astronomical units and the y axis represents the log of mass in Earth masses) to a list of characters as shown in Fig. \ref{fig:encoding}. For the purpose of this work, we use a uniform grid of $N^2$ rectangles with $N=30$ in the logarithm of the semi-major axis $\in [-3, \,3]$ \unit{\astronomicalunit} and the logarithm of the mass $\in [-4, \,4]$ $M_\oplus$ space.
    \item the list of ~25000 planetary system formation models translates in a list of words, which is used to train and test the generative model (see \ref{transformer}). Some examples of such words and corresponding planetary systems are shown on Fig. \ref{fig:real_PS}.
    \item once the generative model has been trained, it can be used to predict new `words'. Each character is then mapped to a synthetic planet, using the inverse mapping as presented above. Since the width of every cell in the grid is non-zero, the exact values of the mass and semi-major axis of the synthetic planets are randomly chosen inside the cell. This inevitably leads to an imprecision in the values of the planet parameters, which can be reduced by increasing the size of the grid.
\end{itemize}

A uniform grid for the encoding was chosen for simplicity purposes. After testing different refinements, $N=30$ seems to be a compromise between capturing the features of the planetary systems and having enough training samples. Note that if the grid is too refined, the generative model might not be able to capture the correlations between the different planets in a planetary system as each character will only be represented, in the training set, a small number of times. On the other hand, if the grid is coarsely refined, several planets might be coded using the same character while having relatively different properties. Lastly, note that there is, inevitably, a loss of information in any similar encoding we would use.

As a straightforward decoding approach to retrieve the planets from a new generated word, we map the character back to its rectangle and use random sampling to compute a semi-major axis and mass. This method can lead to instabilities in the new decoded planetary systems, as a pair of planets (located in the same or two different rectangles) can be stable or instable depending on their precise location in the rectangles. To avoid generating unstable systems, we use a simplistic planetary system stability criterion $\Delta = (a_{p_2}-a_{p_1})/ R_{\text{Hill}} \lesssim 2\sqrt{3}$ \citep{Marchal1982, Chen2024} where the Hill radius, $R_{\text{Hill}}$, is defined as:
\begin{equation}
    R_{\mathrm{Hill}} = \left(\frac{m_{p_1}+m_{p_2}}{3m}\right)^{1/3} \left(\frac{a_{p_1}+a_{p_2}}{2}\right),
\end{equation}
\vspace{0.1cm} and $m_{p_1}$ and $m_{p_2}$ are the masses of two adjacent planets in ascending order of semi-major axis, $a_{p_1}$ and $a_{p_2}$ their semi-major axis and $m$, the mass of the central star.

While it is known that this criterion is an approximate one (in particular, the real dynamical stability of a pair depends on all orbital parameters, not only the semi-major axis), this allows us rapidly infer which planet pairs would be unstable. To decode a generated word into a synthetic planetary system, we sample at random in the rectangles corresponding to the different characters of the word. Then, we check that all planet pairs are stable according to the criterion presented above. If any planet pair is found to be unstable, we attempt at sampling again inside the problematic rectangle, until the pair becomes stable.

It is important to stress the limitations of this encoding-decoding technique. For example, it does not allow to accurately model mean-motion resonances. Additionally, the stability of the system is only evaluated considering pairs of planets and not the global picture. We reserve the development of new encoding and decoding techniques (e.g. to consider mean-motion resonances) as well as better stability criteria to future work.

\subsection{Transformer}
\label{transformer}

Our model architecture is based on the well-known transformer architecture which leverages self-attention mechanisms \citep{vaswani17}. We will not describe in this paper all the details of the transformer architecture, the interested readers can refer to the original paper of \citep{vaswani17}. Our baseline model only comprises of a decoder stack, which is made of three transformer layers, each of them using a single attention head. Contrary to language models, we do not use positional encoding, since the `position' of a planet in the sequence encoding a planetary system is already hinted at by the character representing each planet\footnote{Note that we could have treated planetary systems as unordered sets instead of ordered sequences. Considering them has sequence is useful as, in this case, planet $n+1$ is by construction at a larger semi-major axis than planet $n$. This implies that the conditional probability of planet $n+1$ semi-major axis given planet $n$ is equal to 0 for all semi-major axes smaller than the one of planet $n$. Such property, that would not be true for a set, is learned by the transformer.}. The initial embedding coding for the different characters has a size of 32, while the final feed-forward neural network has one hidden layer with 128 units, and uses the GeLU activation function \citep{GELU}. Our code was developed using PyTorch, following closely the \textit{Makemore} code of A. Karpathy\footnote{see https://github.com/karpathy/makemore}. The total number of parameters of the model is $~60000$. 

\begin{figure}
    \centering
    \includegraphics[width= 0.4\linewidth]{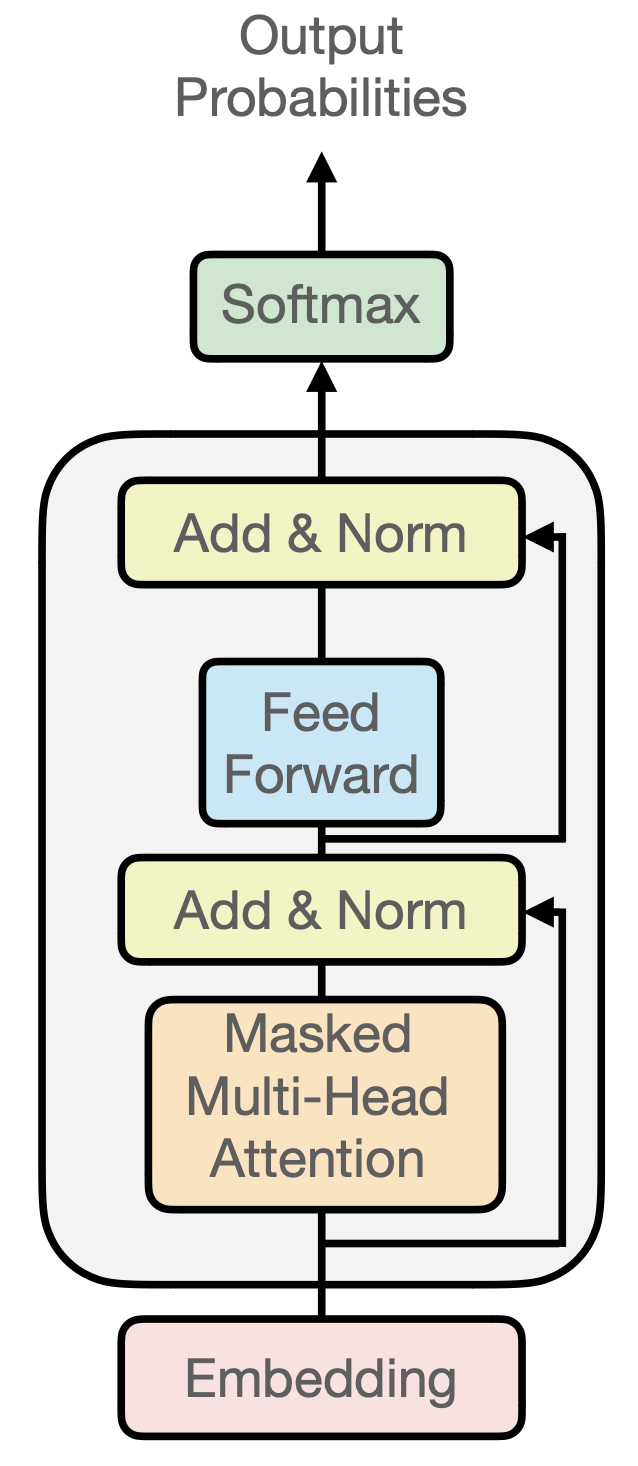}
    \caption{Model architecture. The model is made of an embedding layer (dimension 16 to 128), 2 to 8 blocks (gray rectangle) and a softmax layer which predicts the probability of each character knowing all the previous ones. Each block is made of a masked multi-head attention layer (between 1 and 8 heads) followed by a normalization layer, followed by a feed-forward neural network (one hidden layer, the number of units being four times the size of the embedding size) followed by a second normalization layer. The different tested architectures (number of blocks,  size of the embedding layer,   number of attention heads) as well as the best obtained cross-entropy loss is indicated in Table \ref{table:results}.}
    \label{fig:transformer_architecture}
\end{figure}

\subsection{Training}
\label{training}

Our sample of 25000 planetary system was randomly split into a training set consisting of 24000 samples and a test set consisting of 1000 samples. We trained our model using the AdamW optimizer of PyTorch \footnote{https://pytorch.org/docs/stable/generated/torch.optim.AdamW.html} with default parameters. We chose a learning rate of $5 \times 10^{-4}$ and a weight decay of 0.01. The training was done during 70000 steps where the samples were processed in batches of 32. The entire training set is therefore processed on average every 730 steps. 

For each sample/word, the model is trained to predict the next character knowing the previous ones in the sequence. Each planetary system (or 'word') contains therefore many training sub-samples at once: firstly, we predict the first character, then, we predict the second given the first, then the third given the first two others and so on. 

To evaluate the training phase, the cross entropy loss was used. This choice is justified by its popularity for multi-class classification problems and also because it provides a smooth and differentiable function for optimization \citep{CrossEntropy2023}. In our case, the classes correspond to the different possible characters that can be attributed to a planet.

The training took a few hours on MacBook with M2 Pro chip, and we achieved a best cross-entropy loss of $2.4014$ after ~49000 epochs (one epoch corresponding to the evaluation of one 32-batch of samples). Table \ref{table:results} shows the best test losses obtained for different architectures, and Fig. \ref{training} shows the evolution of the loss as a function of the training epoch for the best model. While a training and test loss which decrease and stagnate over the training period is already a good indication that the model learned the relations between the different characters and how the words were formed, the loss value also provides insights that the training was successful. If a word would be generated randomly from the different possible characters, a cross entropy loss of $log \,N$, where $N$ is the number of possible characters, would be expected. As in our case, we can select from 440 different characters\footnote{One would expect 900 classes from the $N^2$ grid with $N=30$, but in fact, not all characters/rectangles of the grid appear in words, thus do not participate in the cross entropy loss and we get our classes reduced to 440 instead.}, meaning that the cross entropy loss for the worst generative model would be $\sim 6.09$. A cross entropy loss of 0 would mean a perfect model, and a model around 2.40 as in our case, indicates a good model, capable of capturing the essence of the word generation, i.e, the planetary system's architecture.

\begin{figure}
    \centering
    \includegraphics[width=0.9\linewidth]{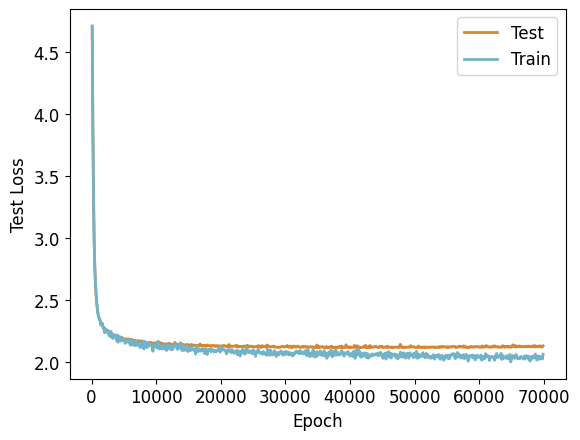}
    \caption{Evolution of training and test cross entropy loss as a function of the training epoch for the best generative model (see Table \ref{table:results}).}
    \label{fig:train}
\end{figure}

We also explored the effect of varying the parameters of our model, namely the dimension of the embedding, the number of blocks, the number of attention heads, and the inclusion of positional embedding (whose size is equal to the embedding). For all the tests, the number of units in the feed-forward network is equal to 4 times the embedding size. Table \ref{table:results} shows that the value achieved for the best loss does not strongly vary for all the models we considered. For the rest of the paper, we will only consider our nominal model.

\begin{table*}
    \caption{Best cross entropy test loss obtained for different architectures of the generative model.}
    \label{table:results}
    \centering
    \begin{tabular}{ccccc}
        \hline
         embedding  & blocks & attention heads & positional embedding & best loss \\
        \hline
  64 & 4 & 4 & no & 2.422051429748535 \\
  64 & 4 & 4 & yes  & 2.445077657699585 \\
  128 & 6 & 8 & no & 2.4423086643218994 \\
  128 & 6 & 8 & yes & 2.429002523422241 \\ 
  32 & 3 & 2 & yes & 2.412593126296997  \\
  32 & 3 & 2 & no & 2.402848482131958 \\
  32 & 2 & 2 & no & 2.4102702140808105 \\
  64 & 3 & 2 & no & 2.4219555854797363 \\
  16 & 3 & 2 & no & 2.425621509552002 \\
  32 & 3 & 4 & no & 2.402952194213867 \\
  \textbf{32} & \textbf{3} & \textbf{1} &\textbf{no} & \textbf{2.40149188041687} \\
        \hline        
    \end{tabular}
    \tablefoot{Our nominal and best model is indicated in bold face.}
\end{table*}
\section{Results}
\label{results}

\subsection{Visual comparison}

We first looked at the mass \textit{versus} semi-major axis diagram of generated planetary systems, trying to see if they can be recognized from the ones resulting from direct numerical simulations. Some examples are given in Fig. \ref{fig:PS}. In some cases, the generated nature of an example can be supposed, while in some cases, it is hard to infer from which population the sample is drawn from.

\subsection{Statistical comparison}

\begin{figure*}
    \centering
    \includegraphics[width=0.95\textwidth]{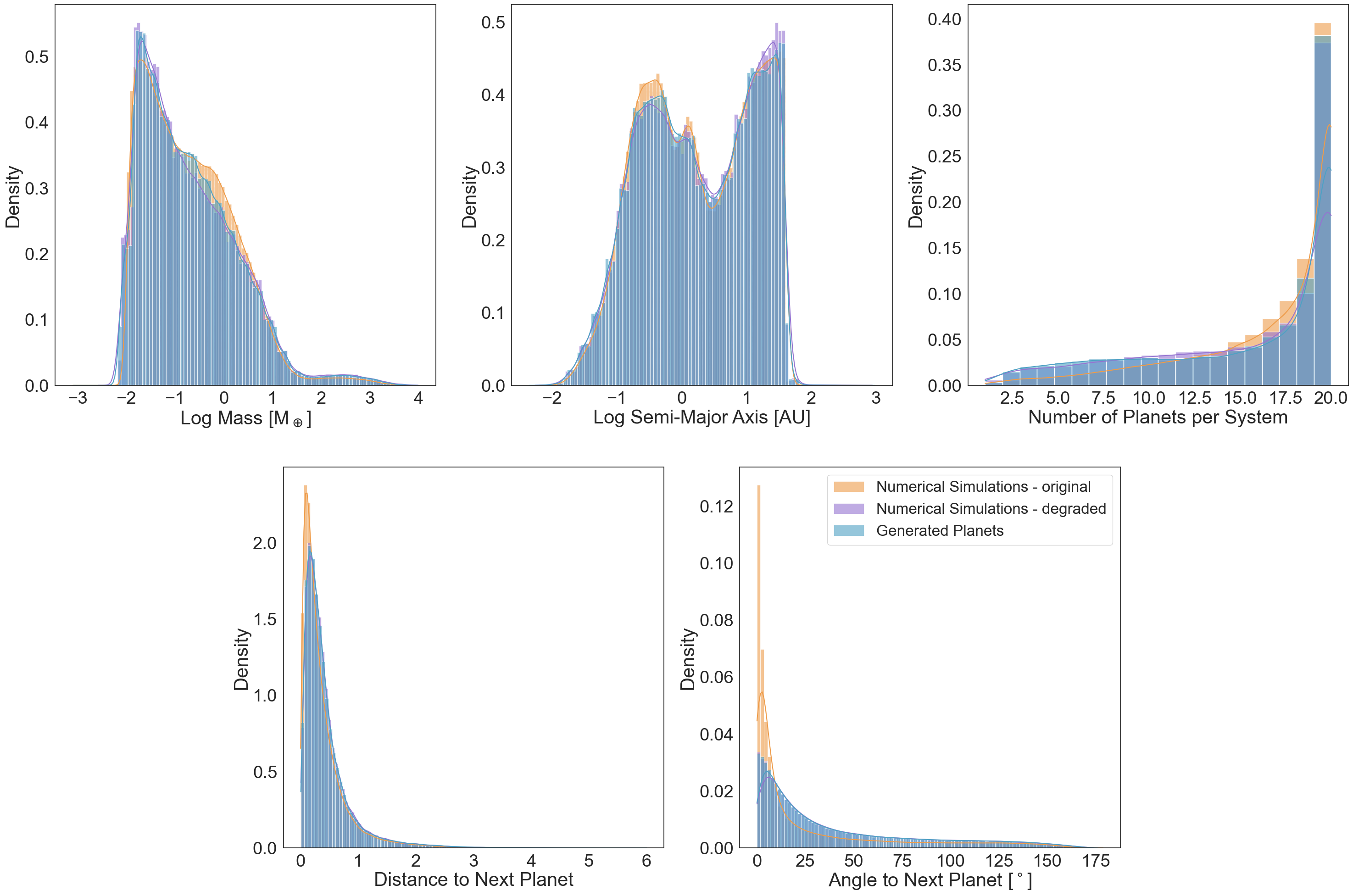}
    \caption{Comparison between different properties of the populations considered. The numerical one shown in orange, the degraded one - encoded and decoded to assess the information loss - is shown in purple, and the one generated by the transformer is shown in blue.}
    \label{fig:comparison}
\end{figure*}

Secondly, we made an initial comparison between the three populations: the original synthetic population (computed using the Bern model), a degraded synthetic population (using the Bern model, but applying an encoding followed by a decoding operation) and the generated population (computed with our generative model). In this case, the degraded synthetic population allows a fair evaluation of the generative model only, while when comparing with the original one, we can evaluate the performances of the whole process (including the encoding-decoding), and assess the total information loss. Fig. \ref{fig:comparison} shows the distributions of various quantities for the two populations. Some quantities are linked to the physical properties of planets (mass and semi-major axis), while some quantities are linked to the presentation of planetary systems as broken lines in a 2D plane. These quantities are:
\begin{itemize}
    \item the distribution of planetary masses (first row, left in Fig. \ref{fig:comparison})
    \item the distribution of planetary semi-major axis (first row, middle)
    \item the number of planets in the system (first row, right)
    \item the (Euclidean) distance between two points representing two consecutive planets in the same system (second row, left)
    \item the angle between two segments of the broken line (second row, right)
\end{itemize}

As can be seen on Fig. \ref{fig:comparison}, the statistical properties of the three populations are very similar in most cases. For the angle and length distribution, we can see that there is a mismatch around the smaller values. This is an artifact from the grid-based encoding/decoding method.

\subsection{Machine learning-based comparison}

Using our nominal model, we produced ~25000 generated systems. Together with the ~25000 synthetic systems, we randomly mixed the two populations, synthetic systems being assigned the label 1, whereas generated ones are assigned the label 0. We then considered 18 different classifiers trying to classify the samples in the mixed population. For the classification problem, the 50000 samples were split in 40000 train samples, 5000 test samples and 5000 validation samples. Our goal was to check if ML-based classification algorithms could distinguish between the two populations. We ran this exercise at different steps (each 100 epochs) of the training of our nominal model, as we expect that very early during the training, it should be straightforward to distinguish between the numerical and generated populations, whereas, in the ideal case, when the generative model is fully trained, it should become impossible to distinguish between the two populations.

Each of the classifiers, after training, produces a single value (between 0 and 1) that can be used to decide if a sample belongs to the numerical population or the generated one. For each of classifiers and each of the training steps of our generative model, we computed the area under the curve (AUC in the following) of the so called receiver operating characteristic curve (ROC curve)\footnote{This curve shows the true positive rate as a function of the false positive rate for all values of the classification threshold (the threshold beyond which we decide a sample belongs to the numerical population).}. A perfect classifier would have an AUC equal to 1, whereas in principle a value of 0.5 indicates a purely random choice. We used the Sci-kit library to compute the AUC\footnote{https://scikit-learn.org/stable/modules/generated/sklearn.metrics.auc.html}, the classifiers were implemented using the Sci-kit (linear, Support Vector, Random Forrest) or Keras (Feed forward deep neural networks) libraries \footnote{https://scikit-learn.org/stable/ and https://keras.io}.

\begin{table*}
    \centering
    \caption{Classifiers considered for machine learning-based comparison with some of their parameters.}
    \label{table:classifier}
    \begin{tabular}{cccccccc}
        \hline
         ID & type  & \#  estimator & max. depth & \#  layers & units per layer & AUC  & AUC  \\
          &   &  &  &  &  & single pop. & Transformer \\
        \hline
  LC & Linear & & & & & 0.57 & 0.54\\
  SVC & Support Vector Classifier & & & & & 0.55 & 0.54\\
  RF1 & Random Forrest & 10 & 5 & & & 0.57& 0.55\\
  RF2 & Random Forrest & 10 & 20 & & & 0.53&  0.55\\
  RF3 & Random Forrest & 50 & 5 & & & 0.55& 0.54\\
  RF4 & Random Forrest & 50 & 20 & & & 0.53& 0.57\\
  DNN1 & Deep Neural Network & & & 1 & 16 & 0.58 & 0.57 \\
  DNN2 &Deep Neural Network & & & 1 & 32 & 0.59  & 0.58\\
  DNN3 &Deep Neural Network & & & 1 & 64 & 0.59 & 0.52\\
  DNN4 &Deep Neural Network & & & 1 & 128 & 0.60 & 0.56\\
  DNN5 &Deep Neural Network & & & 3 & 16 & 0.60 & 0.57\\
  DNN6 &Deep Neural Network & & & 3 & 32 & 0.62 & 0.58\\
  DNN7 &Deep Neural Network & & & 3 & 64 & 0.59 & 0.58\\
  DNN8 &Deep Neural Network & & & 3 & 128 & 0.58 & 0.57\\
  DNN9 &Deep Neural Network & & & 5 & 16 & 0.60 & 0.58\\
  DNN10 &Deep Neural Network & & & 5 & 32 & 0.58 & 0.56\\
  DNN11 &Deep Neural Network & & & 5 & 64 & 0.59 & 0.59\\
  DNN12 &Deep Neural Network & & & 5 & 128 & 0.59  & 0.57\\
        \hline       
    \end{tabular}
    \tablefoot{The AUC is the Area Under the Curve obtained when classifying two similar parts of the same population (see text for details). All classifier were computed using the Sci-kit library (see https://scikit-learn.org/stable/) or (for deep neural networks - DNN) the Keras library (see https://keras.io). AUC Single pop. refers to the AUC when classifying the numerical population only (with random labels), while AUC Transformer refers to the AUC for the last iteration of the generative model (baseline model).}
\end{table*}

Table \ref{table:classifier} summarizes the different classifiers we considered, as well as some of their characteristics. We also quote, in this table, the AUC obtained by randomly splitting the numerical population equally in two parts, one half being assigned the label 0, the other half the label 1. In principle, since this is an homogeneous population, we expect that no classifier can solve this problem. We indeed can see in the table that the AUC for each classifier is close to 0.5. We finally quote in the table the AUC obtained when classifying the numerical population \textit{versus} the generated population computed using our generative model after training. 

\begin{figure}
    \centering
    \includegraphics[width= 0.8\linewidth]{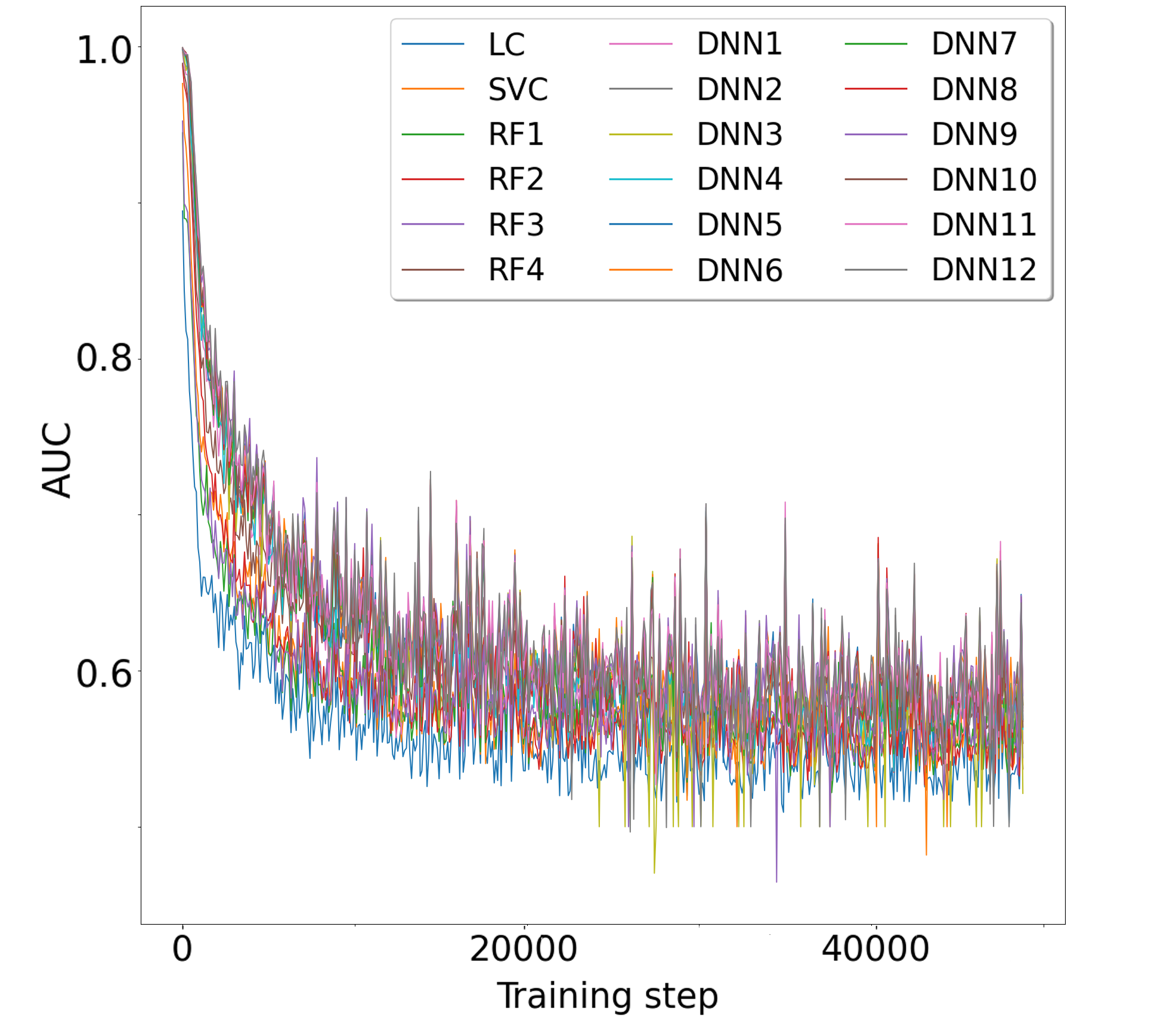}
    \caption{Area Under the Curve (AUC) as a function of the training epoch of our generative model for different classification algorithms.}
    \label{fig:AUC}
\end{figure}

In Fig. \ref{fig:AUC}, we plot the classification performances (quantified by the AUC) of a set of machine-learning classifiers, as a function of the training of the generative model. As expected, in the beginning when there is no or little training of the generative model, all classifiers are able to distinguish between the numerical and generated populations (all values are close to 1). As we train the generative model, it learns how to generate the planetary systems properly (statistically similar to the training sample) and it becomes more and more difficult to distinguish generated systems from numerically computed ones, thus the performances of the classifiers decrease. At the end of the training, the AUC is very close to the values in Table \ref{table:classifier}, showing that no classifier can efficiently distinguish between the two populations.

While it is not excluded that other classifiers could achieve better performances, this gives us confidence that the statistical properties of the generated and numerical populations are very close to each other.

\section{Example}

As an demonstration of a use case of our generative model, we considered the TOI-469 system \citep{Damasso2023}. This system with 3 observed planets, was firstly discovered by TESS with the detection of planet b and later observed using ESPRESSO where two other planets, c and d were found. With further observations \citep{JoAnn2024}, it was possible to characterize with a high precision the properties of the three observed planets, some of these properties being listed in Table \ref{table:toi469_properties}.

\renewcommand{\arraystretch}{1.3}
\begin{table*}
\caption{Properties of planets b, c and d from TOI-469 system.}
\label{table:toi469_properties}
\centering
\begin{tabular}{cccc}
\hline
          & Semi-major Axis {[}au{]} & Mass {[}M$_\oplus${]}  & RV semi-amplitude K {[}m/s{]} \\ \hline
TOI-469 b & 0.1110 $\pm$ 0.0020      & 9.10$^{+0.82}_{-0.79}$ & 2.63$^{+0.20}_{-0.19}$        \\
TOI-469 c & 0.04518 $\pm$ 0.00082    & 4.50$\pm$ 0.32         & 2.04$\pm$ 0.11                \\
TOI-469 d & 0.0673 $\pm$ 0.0012      & 5.14$\pm$ 0.41         & 1.91$\pm$ 0.12                \\ \hline
\end{tabular}
\tablefoot{RV stands for radial velocity.}
\end{table*}
\renewcommand{\arraystretch}{1}

We consider a scenario that could have been followed just after the TESS discovery, when only TOI-469 b was known. We use our generative model to estimate the potential properties and number of planets in the system, knowing only the properties of the b planet. For this, we generate 300,000 planetary systems and select the ones with a planet similar to TOI-469 b, allowing a 10\% uncertainty on the mass and 1\% uncertainty on the semi-major axis. We assume a minimum threshold of detectability of radial velocity (RV) semi-amplitude $K=1.5$ m/s, and we only consider planets which fulfilled this criterion for plotting. From the $\sim$ 300,000 planetary systems, only around 350 have a TOI-469 b like planet.

\begin{figure*}
    \centering
    \includegraphics[width=0.73\linewidth]{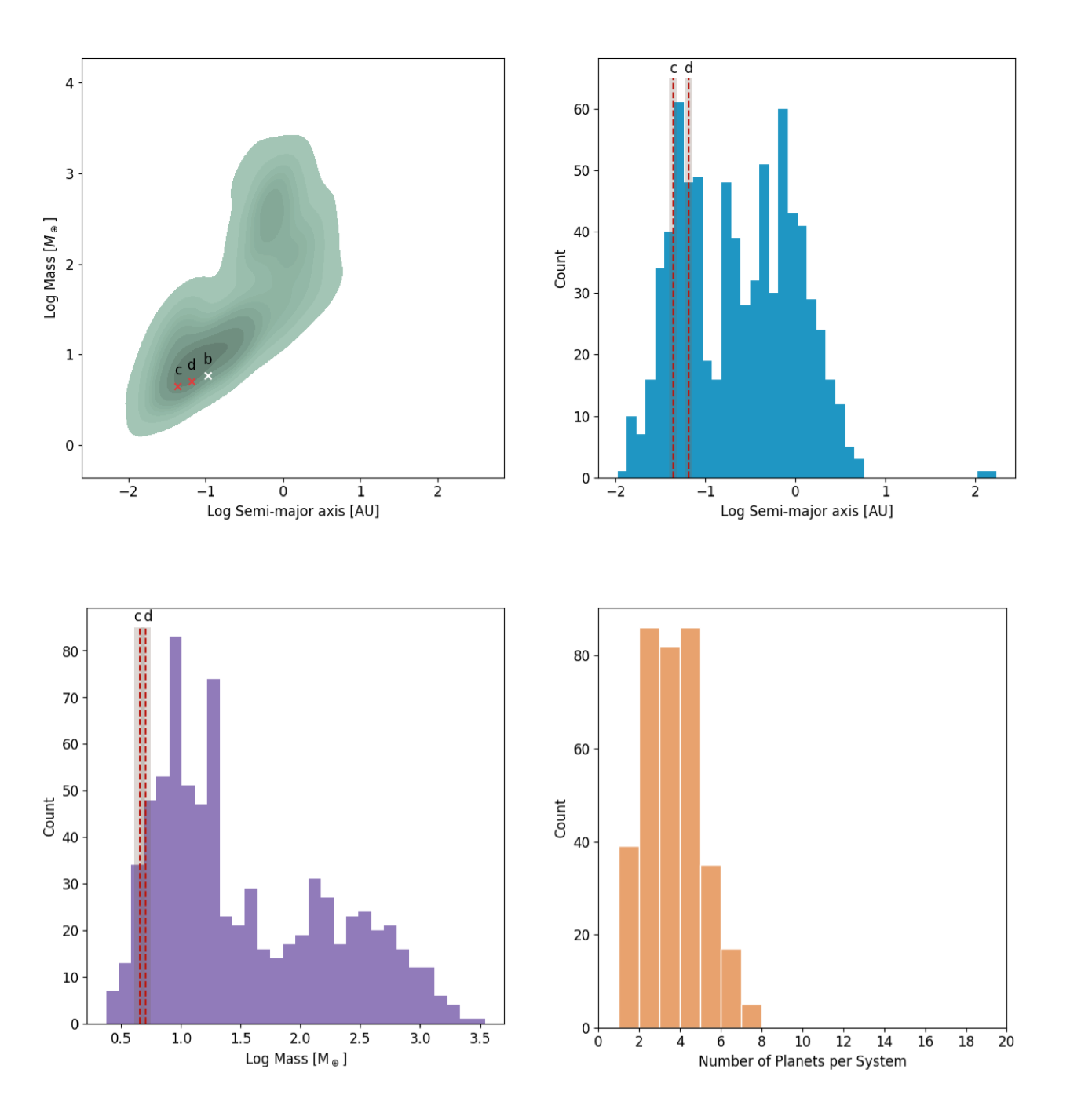}
    \caption{Predictions, based on our generative model, of the properties of planets in systems that harbor one planet similar to TOI-469 b. Upper-left: 2D histogram in log semi-major axis \textit{versus} log mass. The positions of planets c and d are indicated with a red cross, whereas the position of TOI-469 b is indicated by a white cross. Upper-right and lower-left: distribution of log semi-major axis and log mass respectively. The positions of planets c and d is given by the read lines, including a 10\% uncertainty (light red region). Lower-right: distribution of the number of observable planets with $K \geq 0.15 m/s$}
    \label{fig:toi-469-results}
\end{figure*}

The final results can be seen in Fig. \ref{fig:toi-469-results}.The 2D histogram shows two concentrations of planets: one around the position of TOI-469 c and TOI-469 d, and one for giant planets around 1 AU. Considering  the distribution of semi-major axis, planets c and d are located close to the main peak in the histogram. They are, on they other hand, slightly less massive than the peak of the mass histogram. According to our model, the number of planets with a RV semi-amplitude larger than the  adopted cutoff ranges from 2-4, and only 40\% of the systems would have a giant planet (more massive than 100 $M_\oplus$ - such a planet would have been detected easily by ESPRESSO). We therefore conclude that the properties of systems compatible with the properties of TOI-469 b are consistent with the properties of the system characterized by ESPRESSO and CHEOPS.

\section{Conclusions}
\label{conclusion}

\begin{figure*}
    \centering
    \includegraphics[width=0.95\linewidth]{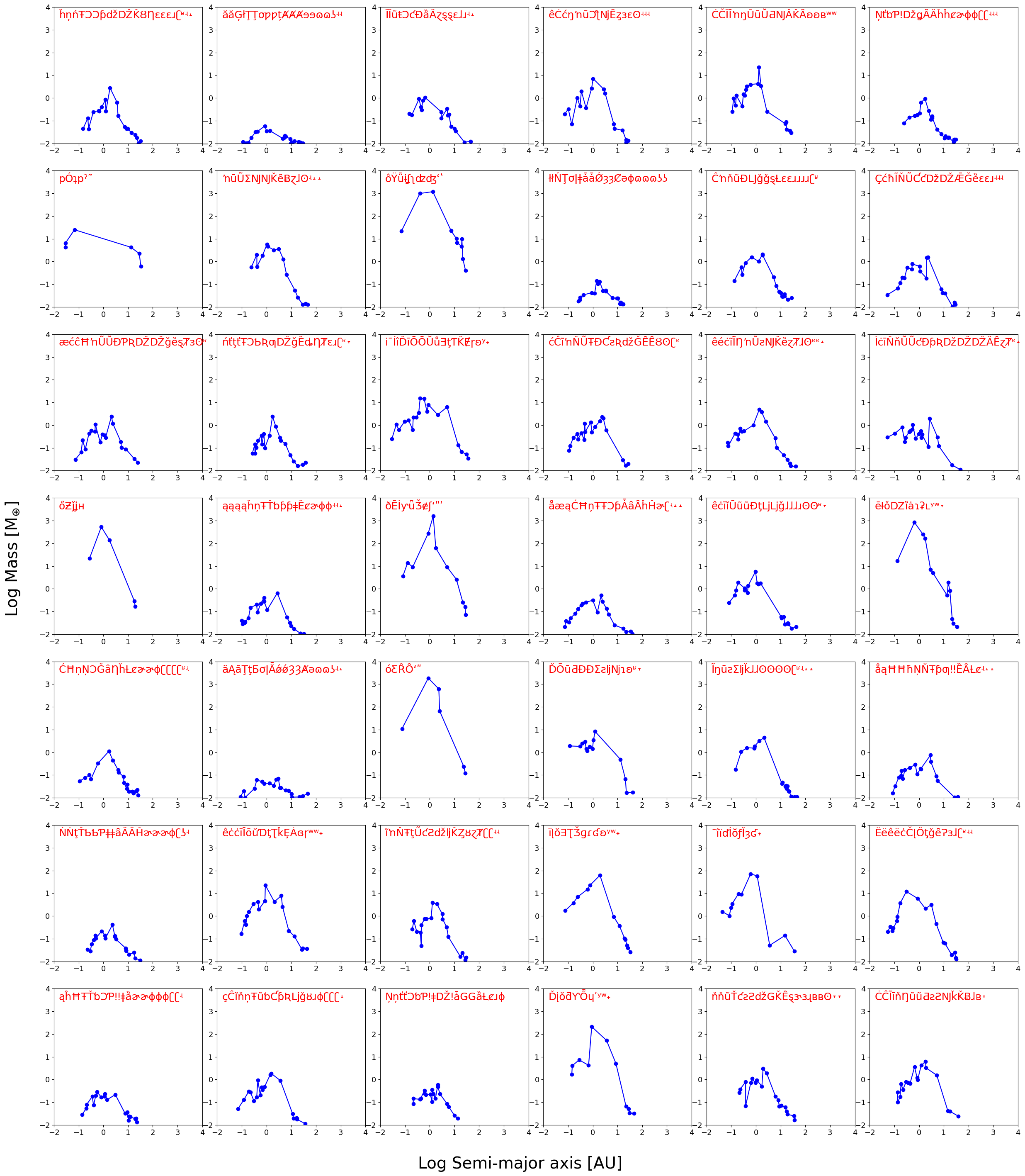}

    \caption{Example of planetary systems, half of them being generated by our generative model, the other half coming from our training dataset. The x axis represents the log of semi-major axis (in AU), the y axis represents the log of mass (in Earth masses). Each planetary system is represented as a broken line joining points, themselves representing the planets. The characters in red in each of the panels is encoding of the planetary system (see Sect. \ref{encoding}). The exact populations, generated or from numerical simulations, to which each system belongs is indicated in a footnote on the Conclusions (Sect. \ref{conclusion}).}
    \label{fig:PS}
\end{figure*}

We developed a model that can be used to compute generated planetary systems. The model, based on the transformer architecture, was trained on a database of ~25000 synthetic planetary systems themselves computed using the `Bern model' \citep{NGPPS1,NGPPS2}, assuming a central star similar to the Sun. We only considered two features of synthetic planets: their total mass, and their semi-major axis, and showed that our generative model is able to produce generated planetary systems whose properties are statistically similar to the ones of the training database. This similarity was first checked first visually\footnote{In Fig. \ref{fig:PS}, generated systems are located on columns 1, 3 and 5.}, then showed by comparing different statistics of planetary systems, and further confirmed using machine-learning algorithms of different types.

Our model can be used to study fine statistics related to planetary system architecture, as well as for computing the probability distribution of some planet's properties, given the observed properties of other planets in the same system. In a future step, we plan to develop this model to condition the generation of sequences on the main properties of the star and the protoplanetary disk in which the system was formed.

\begin{acknowledgements} 
The authors acknowledge support from the Swiss NCCR PlanetS and the Swiss National Science Foundation. This work has been carried out within the framework of the NCCR PlanetS supported by the Swiss National Science Foundation under grants 51NF40$_{}$182901 and 51NF40$_{}$205606. Part of this work has been developed in the framework of the Certificate for Advanced Studies 'Advanced Machine Learning' of the University of Bern. The authors thanks the teachers and students of the CAS-AML. 
\end{acknowledgements}
\bibliographystyle{aa}
\bibliography{ref_transformers.bib}

\end{document}